\documentclass[sigconf]{acmart}
\AtBeginDocument{%
  \providecommand\BibTeX{{%
    \normalfont B\kern-0.5em{\scshape i\kern-0.25em b}\kern-0.8em\TeX}}}

\setcopyright{acmcopyright}
\copyrightyear{2018}
\acmYear{2018}
\acmDOI{XXXXXXX.XXXXXXX}

\acmConference[Conference acronym 'XX]{Make sure to enter the correct
  conference title from your rights confirmation emai}{June 03--05,
  2018}{Woodstock, NY}
\acmPrice{15.00}
\acmISBN{978-1-4503-XXXX-X/18/06}

\usepackage{xcolor}
\usepackage{subcaption}
\usepackage{microtype}
\usepackage{array}
\usepackage{enumitem}
\usepackage{colortbl}
\usepackage{graphicx}
\usepackage{color}
\usepackage{verbatim}
\usepackage{multirow}
\usepackage{balance} 
\newcolumntype{\$}{>{\global\let\currentrowstyle\relax}}
\newcolumntype{^}{>{\currentrowstyle}}

\def\markup{0}
\if\markup1

\else

\newcommand{\st}[1]{}
\fi

\copyrightyear{2023}
\acmYear{2023}
\setcopyright{acmlicensed}
\acmConference[MM '23] {Proceedings of the 31st ACM International Conference on Multimedia}{October 29--November 3, 2023}{Ottawa, ON, Canada.}
\acmBooktitle{Proceedings of the 31st ACM International Conference on Multimedia (MM '23), October 29--November 3, 2023, Ottawa, ON, Canada}
\acmPrice{15.00}
\acmISBN{979-8-4007-0108-5/23/10}
\acmDOI{10.1145/3581783.3612438}

\begin{document}

\title[Designing Loving-Kindness Meditation in VR]{Designing Loving-Kindness Meditation in Virtual Reality for Long-Distance Romantic Relationships}

\author{Xian Wang}
\authornote{Both authors contributed equally to this research.}
\email{xian.wang@connect.ust.hk}
\author{Xiaoyu Mo}
\authornotemark[1]
\affiliation{%
  \institution{The Hong Kong University of Science and Technology}
  \city{Hong Kong SAR}
  \country{China}}
  \email{xmoac@connect.ust.hk}

\author{Lik-Hang Lee}
\affiliation{%
  \institution{The Hong Kong Polytechnic University}
  \city{Hong Kong SAR}
  \country{China}
}
\email{lik-hang.lee@polyu.edu.hk}

\author{Xiaoying Wei}
\affiliation{%
  \institution{The Hong Kong University of Science and Technology}
  \city{Hong Kong SAR}
  \country{China}
}
\email{xweias@connect.ust.hk}

\author{Xiaofu Jin}
\affiliation{%
  \institution{The Hong Kong University of Science and Technology}
  \city{Hong Kong SAR}
  \country{China}
}
\email{xjinao@connect.ust.hk}

\author{Mingming Fan}
\authornotemark[2]
\affiliation{%
  \institution{The Hong Kong University of Science and Technology (Guangzhou)}
  \city{Guangzhou}
  \country{China}
}
\affiliation{%
  \institution{The Hong Kong University of Science and Technology}
  \city{Hong Kong SAR}
  \country{China}
}
\email{mingmingfan@ust.hk}

\author{Pan Hui}
\authornote{corresponding authors}
\affiliation{%
  \institution{The Hong Kong University of Science and Technology (Guangzhou)}
  \city{Guangzhou}
  \country{China}
}
\affiliation{%
  \institution{The Hong Kong University of Science and Technology}
  \city{Hong Kong SAR}
  \country{China}
}
\affiliation{%
  \institution{University of Helsinki}
  \city{Helsinki}
  \country{Finland}
}
\email{panhui@ust.hk}

\renewcommand{\shortauthors}{Wang and Mo, et al.}

\begin{abstract}

Loving-kindness meditation (LKM) is used in clinical psychology for couples' relationship therapy, but physical isolation can make the relationship more strained and inaccessible to LKM. Virtual reality (VR) can provide immersive LKM activities for long-distance couples. However, no suitable commercial VR applications for couples exist to engage in LKM activities of long-distance. This paper organized a series of workshops with couples to build a prototype of a couple-preferred LKM app. Through analysis of participants' design works and semi-structured interviews, we derived design considerations for such VR apps and created a prototype for couples to experience. We conducted a study with couples to understand their experiences of performing LKM using the VR prototype and a traditional video conferencing tool. Results show that LKM session utilizing both tools has a positive effect on the intimate relationship and the VR prototype is a more preferable tool for long-term use. We believe our experience can inform future researchers.

\end{abstract}



\maketitle

\section{Introduction}

Meditation has been shown to provide numerous specific health benefits, such as anxiety reduction, 
and meditation-based interventions are growing in popularity in the fields of psychology and healthcare~\cite{behan2020benefits,lippelt2014focused}. \textit{Loving-kindness meditation (LKM)} has been shown to be an effective intervention in clinical psychology for fostering compassion and empathy, which are positively associated with the experience of romantic relationships~\cite{atkinson2013mindfulness, gencc2021transforming, cunningham2012loving, mapp2013relationship}, and many studies have used LKM for couples therapy, such as after infidelity healing~\cite{cunningham2012loving}, enhancing couples' relationships~\cite{carson2006mindfulness,bohy2019mindfulness,atkinson2013mindfulness}. In the clinical situation, LKM is the basis for therapy methods, for example, Mindfulness-Based Relationship Enhancement (MBRE), consisting of loving-kindness meditation, partner yoga, mindful touch practice, and eye-gazing practice, etc.~\cite{carson2006mindfulness,kabat1990full,atkinson2013mindfulness}.
Numerous individuals find themselves in long-distance relationships due to career or academic opportunities and unforeseen events. 
Isolating from the romantic partner tends to intensify relationship issues such as incorrect speculation, separation protest, abnormal attachment styles like attachment anxiety, etc.~\cite{bowlby1969attachment, singh_1988, bowlby1977making,bowlby1973attachment, bowlby2008loss, carole2010long}, due to lack of physical contact, insufficient sensed presence, the latency of synchronization of information and compassion, and so on~\cite{kjeldskov2004using, greenberg2009awareness, bales2011couplevibe}. 

Due to the limitation of physical location, people cannot attend face-to-face loving-kindness meditation sessions in the same physical space. An increasing number of meditation
programs are beginning to be conducted via online video conferencing (VC)~\cite{meissner2017video,walumbe2021pain}. 
However, VC cannot provide a strong sense of connection and presence. Additionally, touch practice cannot be performed when using VC as it only provides images and sound. 
In comparison, virtual reality (VR) can provide a more immersive 
environment for people in long-distance relationships to conduct LKM~\cite{mcveigh2019shaping, mcveigh2018s,maloney2021social}. 
Inspired by these pioneering efforts, we hypothesized that VR could provide assistance to couples in long-distance relationships for LKM and may offer new opportunities to promote healthy long-distance relationships. Thus, we aim to explore the following open research question (\textbf{RQ}): 
\textit{How best to design VR elements (e.g., virtual environments and interactive objects) to support long-distance romantic partners performing LKM?}

Our research aimed to answer the RQ using a participatory design workshop and prototype evaluation. We initially conducted workshops with 7 pairs of romantic partners, who experimented with LKM and experienced various VR meditation environments, then designed VR environments and elements that they would love to have when performing LKM together. Interviews revealed preferences for secure, restorative environments featuring life-related elements and realistic avatars; necessity of audio communication; and preference for physical interactions such as hand-holding, hugging, etc. There was lesser inclination towards visualizing bio-signals and privacy. These findings shaped the design of a VR prototype that was subsequently tested with eight couples (N=16) performing LKM via both the prototype and traditional VC methods. Results demonstrated that both approaches enhanced relationships by boosting positive affect and self/partner compassion, while reducing attachment avoidance. The VR prototype, however, additionally decreased negative affect and attachment anxiety, significantly improving partner compassion compared to traditional methods.


In sum, we make the following contributions: (1) Through participatory design workshops with people in long-distance romantic relationships, we derived design considerations for creating LKM experiences in VR. (2) We designed a VR prototype for romantic partners to perform LKM based on the design considerations and evaluated its viability to inducing positive experiences in comparison with VC.

\section{Background and Related Work}

\textbf{Loving-kindness meditation (LKM)}.
LKM aims at cultivating acceptance and compassion towards oneself and others, including known and unknown humans, living creatures, and even everything in the world ~\cite{zeng2015effect}. To elicit positive thoughts like acceptance, gratitude, joyfulness, and love, practitioners recite phrases like “may X be happy/joyful/peace; may X be away from suffering” in a low voice or silently in their minds ~\cite{graser2018compassion}. 
LKM is crucial to couple therapy in clinical psychology ~\cite{gehart2012mindfulness}. For instance, practicing LKM procedures can arouse the \emph{relaxation response}~\cite{benson1974relaxation}, manifested in a calm and relaxed state, both emotionally and physiologically ~\cite{gottman2001marriage}. 

\textbf{Long-distance relationship}. A long-distance relationship is a particular situation between couples referring partners who live and work in different geographical locations for career or education and reunite for face-to-face interaction by periodically travelling~\cite{carole2010long}. 
According to \emph{attachment theory}, approaches and behaviours become restricted to be \emph{available, sensitive, and supportive} in regard to the partner’s needs in the long-distance situation. This leads to unstable emotions, incorrect speculation, separation protest, abnormal attachment styles like attachment anxiety, etc.,  before, during, and after the reunite period ~\cite{bowlby1969attachment, singh_1988, bowlby1977making,bowlby1973attachment, bowlby2008loss, carole2010long}. Specifically, feeling insecure, lonely, grieving, and lacking nonverbal expressions like touch and hug, tend to lead to misunderstanding and arguments.  
Thus, it is worth exploring technology to support LKM for long-distance relationships as it can help couples calm down, raise positive affect, and improve acceptance and compassion.

\textbf{Technology Support for Meditation}. We present two types of technology support for people to perform LKM or meditation. 
First, Video conferencing (VC) via tools like ZOOM and Skype can provide simultaneous auditory and visual interaction ~\cite{dainton2002patterns} and is feasible for people to proceed with meditation sessions. For example, Campo et al. demonstrate the feasibility of using a VC platform to conduct self-compassion interventions for cancer survivors~\cite{campo2017mindful}. 
However, availability of instructors, space issues, camera challenges, and lack of touches are still limited factors ~\cite{muntean2015synchronous}. 

Mobile apps have also been designed to ask with meditation. For instance, Vacca et al. designed an APP based on mobile phones that guide the user to conduct LKM by interactive audio and visualization ~\cite{vacca2016designing}. The visualization allows a more friendly meditation experience for meditation navies. Mah et al. an interactive system for compassion cultivation based on Buddhism inspiration ~\cite{mah2020designing}. However, such apps were mostly designed for a single user to perform meditation. In contrast, we seek to design experiences for couples in long-distance relationships to perform LKM.


Second, Multi-user VR provides a promising solution for our goal for two reasons. Prior work showed that VR could assist meditation practice efficiently. For example, the VR system RelaWorld provides interactive visual guidance for meditation~\cite{kosunen2016relaworld}. The TranScent provides synchronized olfactory sensation besides auditory and visual experience in VR~\cite{lai2021transcent}. ZenVR supports an immersive Zen garden and room environment and mentor in VR~\cite{feinberg2022zenvr}. Moreover, multi-user VR can provide social interaction experiences closer to face-face, especially for intimate relationships where a sense of closeness is strongly needed. Prior work found that social VR could replicate real-life activities, improve co-presence, and allow diverse self-embodiment ~\cite{zamanifard2019togetherness, moustafa2018longitudinal}. Rukangu et al. verified that a shared virtual family room in VR could help maintain a long-distance relationship ~\cite{rukangu2020virtual}. In some novel explorations, haptic technology can realize the kissing behaviour in VR, which is a typical action in an intimate relationship~\cite{zhang2021turing}.

Only a few prior works investigated multi-user VR for LKM or mindfulness exercise. In the DYNECOM prototype~\cite{salminen2019evoking}, users' EEG signals and respiration rhythm are visualized beside their avatars in the virtual environments while doing LKM. Even though the visualization of biosignals improves the sense of social presence, the prototype was not tailored for couples or people in a romantic relationship. For example, the avatar is gender-neutral to avoid any identity information. The JeL~\cite{desnoyers2019jel} presents underwater scenes with two jellyfish as the avatars and one interactive virtual coral. The jellyfish's floating is a visual guide for users to do mindfulness breathing exercises, and the coral's growing offers a creative interaction for two users to maintain synchronized respiration. The interaction strengthens the bond between two users; however, the prototype was not intended for those in romantic relationships. In sum, it remains unknown how couples in long-distance relationships would want VR to be designed to help them perform LKM, for example, how their avatars should be designed, what virtual scenarios they would want to perform LKM in, and what items should be included for them to interact with. In this work, we sought to answer these questions through participatory design workshops and to gather their experiences and feedback on a VR prototype designed based on the workshop findings. 


\section{Participatory Design Workshop}


\subsection{Participants and Procedures}

Participants were recruited through social network advertisements, consisting of students from two local universities. The study required couples with an interest in meditation or design techniques, preferably with long-distance relationship experience. Of the 9 couples who registered, 7 couples (14 participants) aged 22-27 (M=23.64, SD=1.39) participated, all from Eastern cultural backgrounds and having experienced long-distance situations (>1 year).

\begin{figure}[h]
  \centering
  \includegraphics[width=1\linewidth]{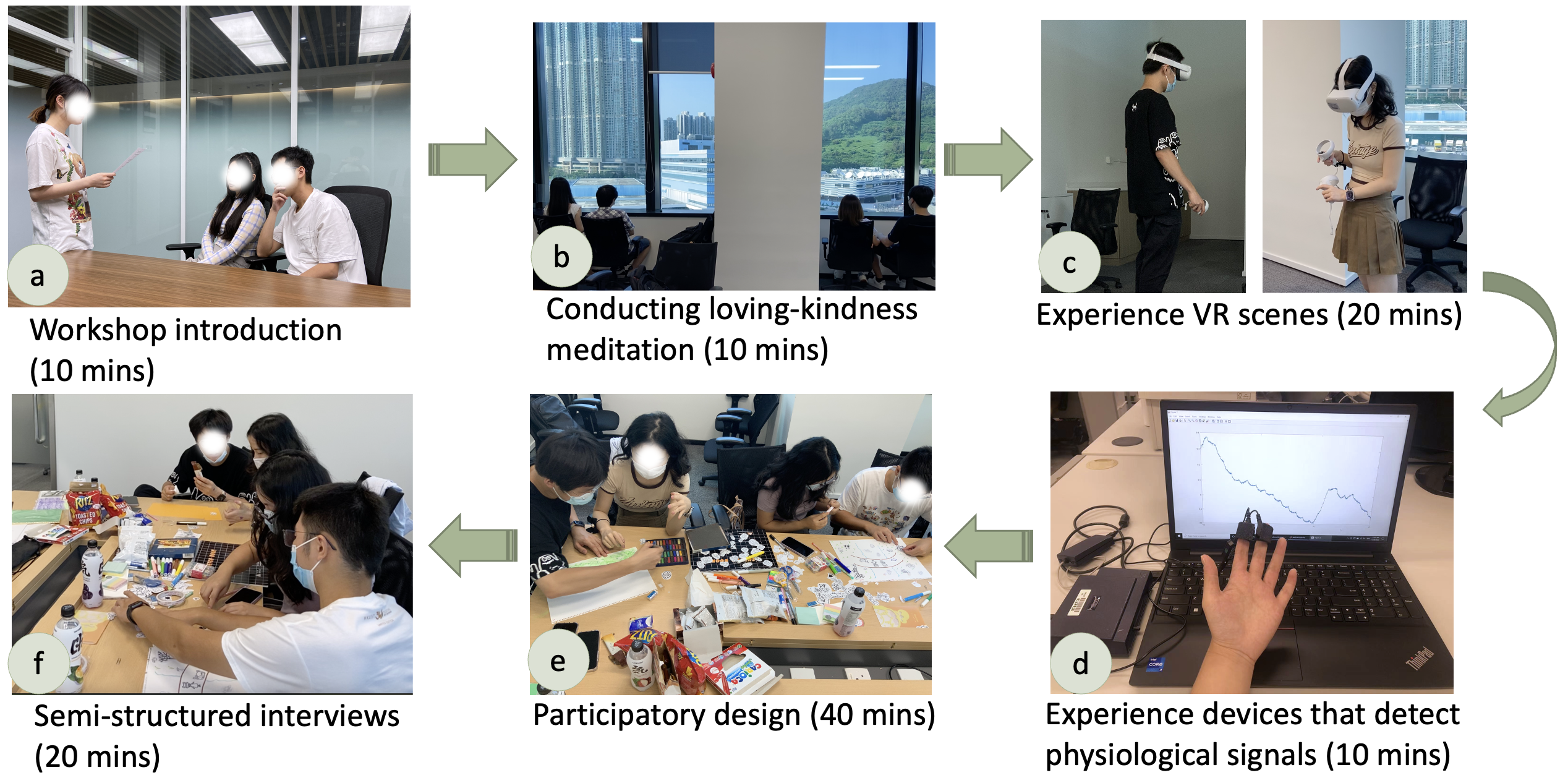}
  \caption{Flow of the workshop. } 
    \Description{Flow of the workshop. } 
  \label{fig:flow}
\end{figure}

Figure~\ref{fig:flow} shows the workshop procedures. The workshop entailed experiencing VR scenes as a couple, with one partner in a separate room to simulate long-distance conditions. Couples maintained voice calls while simultaneously exploring 11 pre-made virtual environments (footnote \#1). After the VR session, participants ranked scene favorability, answered follow-up questions, and underwent a physiological signal detection session~\cite{wang2022reducing}.

Subsequently, couples collaboratively discussed meditation applications for remote partners and sketched designs on paper during a 40-minute group session. The workshop concluded with semi-structured interviews with each couple. All participants provided informed consent and received a \$20 compensation.

\subsection{Materials and equipment}



\subsubsection{Guided audio for LKM}

Beginners often need guidance to enter the meditation state; however, we did not find suitable material in public repositories. 
After consulting with experts, we make a customized guidance script for this workshop, referring to literature on LKM~\cite{sujiva2009meditation, seppala2009loving, vacca2016designing}. The literature suggested that the practice for beginners may contain these essential procedures: finding a comfortable posture, relaxing the physical body and mental state, and proceeding to give wishes to the compassionate objects (oneself, other people, and the world) by concentrating on reciting wishful sentences silently. Therefore, our customized script could be divided into three parts: guides the users to (1) calm down by taking several deep breaths. (2) to concentrate on themselves and give best wishes to themselves through key sentences like: \emph{``may I be safe, may I be happy, may I be good at everything.''} (3) to concentrate on their romantic partners and give best wishes to them in similar sentences: \emph{``may you be safe, may you be happy, may you be good at everything''.}  

To avoid a sense of dissonance, Microsoft Azure~\cite{MicrosoftAzure} automatically transcribes the guided text into audio, rather than the researchers' dubbed voice, with a background sound of Indian yoga meditation background music. All authors, and one invitee with meditation experience, auditioned for the guided audio, which worked well and was uniformly approved by all. We kept the length of the guided audio at about 10 minutes, i.e., 
effective time duration 
for studying the short-term effects of meditation~\cite{seppala2014loving, Ralph2016}. 
\subsubsection{VR scenes}


Our finding in the demographic survey indicated that 71.4\% of the participants did not have VR experiences. Therefore, it is necessary to arrange the VR experience session, i.e., participants learn the capability of VR by mastering the essential operation of the device and experiencing diverse and immersive scenes. T
Vegetation, water, wind, animals and Buddha statues are the virtual environments and objects that previous studies used in VR meditation. We synthesized and designed 11 scenes (see all scenes~\footnote{\url{https://drive.google.com/file/d/1OlJLP_2RVi8DxfO2QSwGAp1Eb1IqYobS/view?usp=share_link}}), covering most of the elements~\cite{wang2022reducing}: 1) Woods with river; 2) Woods with river and cartoon animals; 3) Deciduous woods; 4) Deciduous woods with realistic animals; 5) Bamboo forest; 6) Calm beach; 7) Underwater; 8) Luminous forest; 9) Colorful universe; 10) Mystery temple; 11) Japanese wamuro. 


\subsubsection{Design expression materials}

In addition to markers and sketchbooks for drawing, we also provided cut-out cards of elements such as various clouds, trees, flowers, water, animals, incense burners, etc., for participants to directly paste on the sketch paper. Other stationery includes scissors, watercolor pencils, chalk, transparent adhesive, glue stick, sticky notes, markers, pencils, etc. A laptop is available for participants to search for visual references. 


\subsubsection{Equipment and Venues} A MacBook Air M1 played guided meditation audio for about 10 minutes, while two Oculus Quest 2 presented VR experience sessions. Also, physiological signals were detected by ProComp5 Infiniti basic package with the skin conductance sensor(SC-Flex/Pro)~\cite{ThoughtTechnology}. The measured data were displayed in real-time using MatLab. The study was conducted in an empty and quiet office area of a local university, for the sake of separation and calmness.



\subsection{Analysis and Findings}
All interview sessions were recorded with the participants' permission. 
Also, participants' design drawings, behaviours, and any notes taken by the researchers were discussed and analyzed. Third, participants were invited to score the VR environments based on their favorability with an 11-Point Likert Scale (1 = the least favourite, 11 = the most favourite) after the VR experience. 
Qualitative data proceeded with thematic analysis, referred to~\cite{wienrich2020mind, dollinger2021challenges, terzimehic2019review} and adapted to our case. 
VR is defined as 
\emph{virtual environment, virtual interactive
objects, virtual others, and virtual self-representation}~\cite{wienrich2020mind}. 
We adapted the themes as \emph{virtual environment, avatars (virtual others and virtual self-representation), interaction in VR (virtual interactive objects)} and bio-signals detection. In addition, a \emph{privacy} theme and a \emph{design process} theme were added since intimacy social behaviors were included~\cite{maloney2021social}. The coding process was proceeded separately by two researchers, and any differences were fully discussed until both researchers reached consent. 
Finally, as for the quantitative data, we presented descriptive statistics about participants' favorability for VR scenes to explore or verify participants' preferences. 



\subsubsection{Meditation Virtual Environments}
\label{sec:virtualEnvironments}


 
In terms of virtual environments to mediate in, participants preferred the environments to offer \textit{natural elements}, \textit{sense of security}, \textit{personalized}, and \textit{dynamic} experiences. Next, we elaborate on each of these aspects. 

\textbf{Nature.} 
\label{sec:nature}
All participants drew natural elements in their designs (footnote \#2). Example elements include plants (e.g., grass, flowers, trees, and rivers) and animals (e.g., birds, cats, dogs, fish). 
When asked for their rationales, they expressed a desire to reconnect with nature and escape from stressful work environments such as computers, offices, and motorways in order to relief themselves. In addition, five couples drew animals, such as birds, cats, dogs and fish, in their planned scenarios. The participants' preferences for animals were also in line with how attractive they ranked the VR scenes, see all illustrations in\footnote{\url{https://drive.google.com/file/d/1Z6pvmigNvUtZho5yA8uz5AdXclfvdSL6/view?usp=share_link}}: 1. Woods with river \emph{VS.} 2. Woods with river and cartoon animals, and, 3. Deciduous Woods \emph{VS.} 4. Deciduous Woods with realistic animals ).  \textit{``... There is a cat sleeping on the windowsill, I want to have cute animals in the house, in the real world I do not yet have a cat ...'' - $C1_F$}


\textbf{Sense of security.}
\label{sec:senceOfsecurity}
All participants designed elements and spaces that could provide them with some form of safeguarding and support, ranging from a bridge (footnote \#2~a), an airship (footnote \#2~d), a house (footnote \#2~b, c \& d). 
Moreover, they expressed a desire to sit, stand, or lie down on some form of stable surface in a secure location from which they may see the rest of the surrounding environment. \textit{``...Two floors, wooden spire, small but cozy second floor loft where we can lie down and relax together, with moonlight coming in through the windows... '' - $C5_F$}



 \textbf{Personalization.} Based on participants' design drawings and our interviews, we found that participants would like to stick to one environment as the basestation, where they could do LKM regularly without needing to find a new place every time. With such a stable basestation, they could gradually personalize the items there and create shared memories for themselves when doing LKM.  
\textit{``...We can also arranged this space together, we can decorate the same room online. There would be some furniture models, or maybe, wallpapers, to select, so that we can decorate together.'' - $C7_F$}



\begin{figure}[H]
    \centering
    \includegraphics[width=1\linewidth]{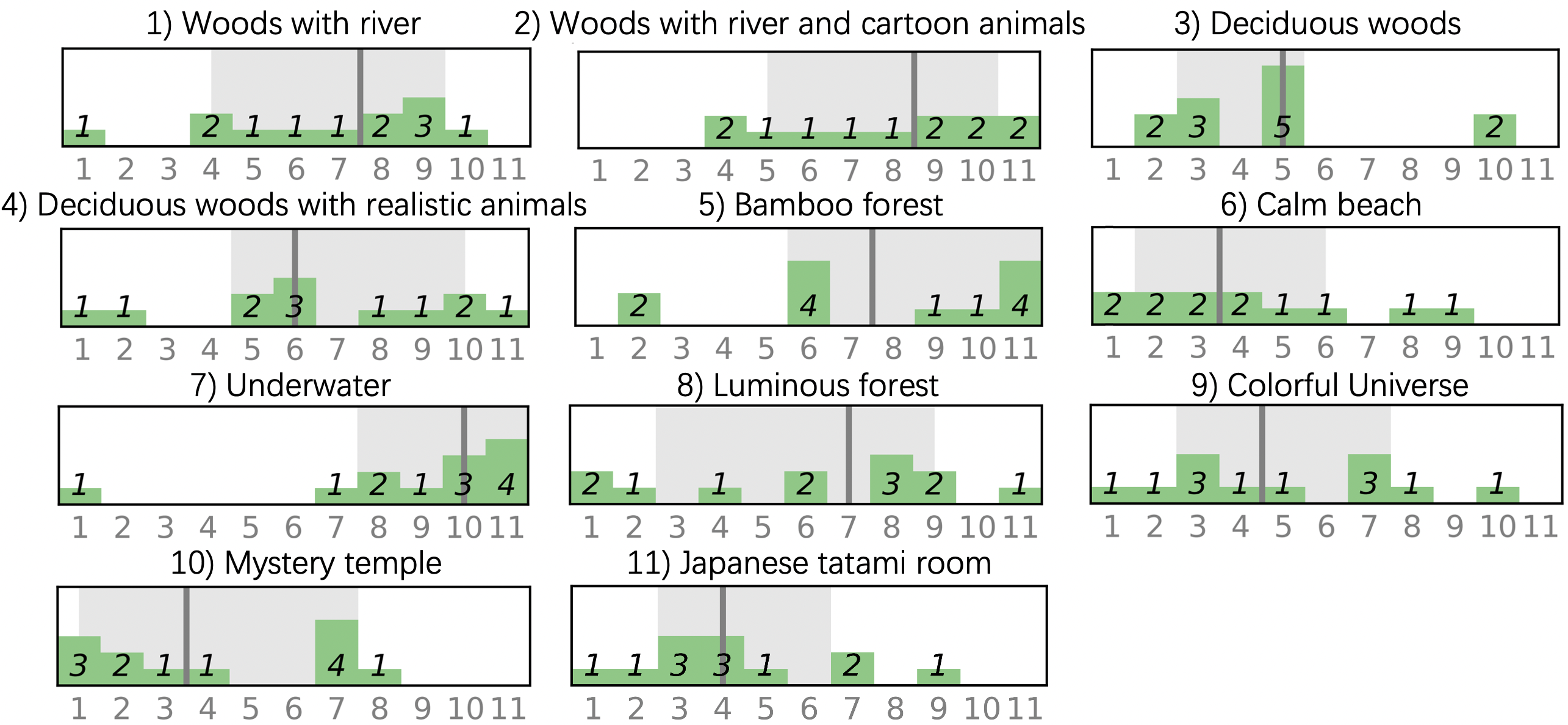}
    \caption{Ranking of Environments Favorability. 
    (Highest: Underwater (7); Lowest: Calm Beach (6)).}
    \Description{Ranking of Environments Favorability. Participants were invited to rank the environments based on their favorability with a 11-Point Likert Scale (1 = the least favorite, 11 = the most favorite). The Underwater environment (7) has the highest mean values; the Calm Beach environment (6) has the lowest.}
    \label{fig:Ranking of Environmental Favorability}
\end{figure}

\textbf{VR Environments Favorability.}
Figure \ref{fig:Ranking of Environmental Favorability} depicts five natural environments (1, 2, 5, 7, 8) receiving relatively better evaluation than the non-natural elements environments (9, 10, 11). 
Three natural environments (3, 4, 6) received relatively lower favourability. Participants indicated that the impermeable space in dense deciduous woods (3, 4) made them feel insecure. Participants ($C5_F$, $C5_M$) said the beach scene was boring and monotonous at first sight, but there seemed to be some unknown threats behind the calm. This finding was consistent with the previous results that people have a preference for nature and demand for security simultaneously (See section~\ref{sec:nature} Nature and Sence of security). 

Environments (9, 10, 11) received a relatively lower ranking. More than half of the participants described that the colourful universe (9) made them feel weird, empty around and sometimes a little bit blurry eyes and dizzy when viewing the dynamic colourful aperture. The only two indoor scenes, namely Mystery temple (10) and Japanese wamuro (11), received the same feedback, such as depression, gray, monotonous, not open and not bright enough. In addition, the decorations, the sculpture and the reliefs on the wall contributed to the mystery and tension.

\subsubsection{Avatars}
\label{sec:avatars}
All participants expressed a strong desire for the avatars to accurately represent themselves, both physically and mentally (e.g., their appearance and mental status), in order to provide each other with presence and a sense of companionship.

All the couples preferred to be represented by human-like avatars, and only one couple accepted to become animals, plants, or other objects, but they emphasized that the default setting should be a humanoid avatar. Their first preference was for avatars that closely resembled real people. 


One couple proposed a ``look-alike'' solution. \textit{``His avatar can not be exactly the same as him, but it needs to be alike, that is, the avatar's specific actions, reactions and expressions or clothing features will make me instantly associate with him, these characteristics are unique to him.'' - $C7_F$}
Half of the participants also stated that they required the avatar to display or express their emotional status. 
Additionally, participants mentioned that the initialization or setting of the avatars could be another potential interactive channel outside the meditation experience, for example, designing or choosing avatars for each other.

\subsubsection{Interaction in VR} We explored the desired interactions in the VR environment in two aspects: (1) between couples; and (2) with the virtual environment.

\textbf{Interaction between couples.} 
\label{sec:fundingInteraction}
All the couples preferred to sit or lie shoulder-to-shoulder in the static position relationship in VR: a) Siting shoulder-to-shoulder on the bridge, b) Sitting on a double sofa), c) Lying shoulder-to-shoulder on a swing, d) Sitting Shoulder-to-shoulder view outside the spaceship), they gave the following reasons: shoulder-to-shoulder sitting feels closer and, more convenient for the interaction between the two of them, while the face-to-face approach makes them feel formal and not relaxed enough, but also that they are not close enough to each other. 
At the same time, they also said that if the scene did not provide other options, they can also sit face-to-face, such as in the previous VR scene in the Japanese room.

Voice communication is a necessary feature, it ensures that couples can talk in a virtual environment, but in addition to the function of voice communication, all couples mentioned some non-verbal interaction between themselves as a transition before meditation, and depending on the design of all couples, these types of interactions could be hand-holding, hugging, head touching, and kissing. Hand-holding was the interaction mentioned by most couples. 


Some of the interactions designed by couples have stronger personal preferences and are highly relevant to the scene. For example, couples who like to play video games wanted to add a shooting game scene in addition to the meditation scene. Some couples (N = 3) preferred simple interactions because they feared that complex interactions would disrupt the meditation effect and prevent them from concentrating on the meditation activity.
However, lying or sitting together to see, feel, and hear the elements of the scene is generally acceptable. 
Many (N = 4) mentioned the haptic sense, and some mentioned (N = 2) the olfactory sense. 

\textbf{Interaction with the virtual environment.} 
All couples who included pets in their scenes wanted the ability to interact with them, such as slow petting and caressing them. 
Couples were interested in adjusting some overall environmental settings (e.g., time, light, and seasons)rather than manipulating specific objects like controlling the growth and morphology of a tree. This is corresponding with the \emph{dynamic environment} in Section ~\ref{sec:virtualEnvironments} .

\subsubsection{Bio-signals detection}
To our surprise, all couples showed relatively less interest in visualization or in-depth interaction of physiological signals in the virtual environment. 
In previous studies \cite{roo2017inner, jarvela2021augmented, desnoyers2019jel}, it was common for people to see or interact with their own physiological signals. 
First, participants were not inclined to the type of real-time data visualization such as a plot, bar, scatter, etc. They raised several concerns. Two pairs said they did not want to affect each other if they were in an anxious or stressed state, as manifested in the data visualization. Another pair expressed that the type of visualization would draw their attention to the result or outcomes, which might cause extra stress and tension. When we further asked the participants how they would like to integrate the physiological signals detection in the VR meditation, they preferred intuitional and artful visualizations such as emoji or heart-shaped patterns. They even said that emojis based on physiological detection could be only one option, and users could choose the one showing their real state, or others.

\subsubsection{Privacy}
When the experimenters asked the couples if they needed to add privacy features to the program, such as certain virtual objects that only an individual could see and not share with each other, the couples said no.
All said they did not need such a feature and that all content could be shared with each other. 
One participant looked at his partner and jokingly said, 
\textit{``Couples don't need privacy.'' - $C3_M$}

\section{Prototype and Evaluation}

\subsection{Prototype Description}
Based on the findings of the participatory design workshop and the literature, we conducted several rounds of discussion to clarify key design considerations of a VR LKM app, as follows. 
\subsubsection{Multi-user VR function.} The prototype was developed in Unity3D. We used Photon Engine to achieve network connection and multi-user function. The final program is packaged as APK files and imported into two Meta Quest 2 as VR head-mounted displays. Therefore, users in different physical locations can log in and enter the same VR space with their VR devices.  

\subsubsection{VR environments.}
Based on the results of the workshop (see Section~\ref{sec:virtualEnvironments}), our prototype realized a home scene: a bedroom with a full wall of the floor-to-ceiling window (see Figure~\ref{fig:demoScense} b). The user can see the nearby garden and the distant sea view through the windows. The bed has a comfortable-looking comforter and pillows. A mirror on the wall allows users to see their avatars. A little cat is lying behind the TV. The bedroom is connected to a small checkroom, where bathrobes and pajamas are hung. We added some white noise and leisurely footsteps to the background sound of the home scene. The scene of a home with a great natural view is the one that best meets our design goal: safety, a sense of belonging, and nature. Besides, the home scene is acceptable to the majority of users. Although the underwater scene received high favorability in the workshop, a few workshop participants mentioned that they felt uncomfortable in this environment because they could not swim. 

\subsubsection{Avatars.} 
\begin{figure}[h]
  \centering
  \includegraphics[width=1\linewidth]{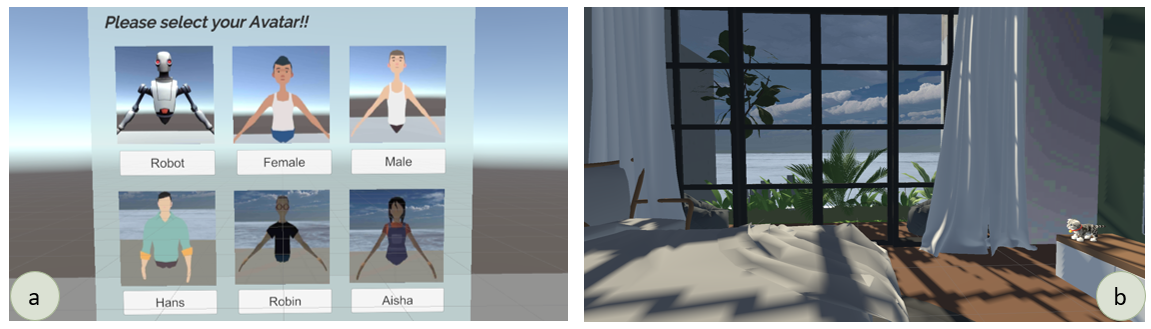}
  \caption{Prototype's interfaces for selecting avatars and 3 meditation scenes. a) 6 avatars selection; b) a home scene.}
  \Description{A screenshot of the prototype's interface for selecting avatars and three meditation scenes. a) 6 avatars selection interface, all of them are humanoid; b) home scene, through the floor-to-ceiling windows users can see the sea and garden. The home has a double bed with a kitten lying on the TV stand at the end of the bed}
  \label{fig:demoScense}
\end{figure}

Our prototype provides six avatars to choose from (see Figure~\ref{fig:demoScense}a). Among them, \textit{Female} and \textit{Male} are the basic avatars that represent different sex. \textit{Robin} and \textit{Aisha} are avatars that can reflect a certain design style in hair and clothing besides sex. \textit{Hans} provides an extra option for males. \textit{Robot} is a nonsexual option. In DYNECOM, the avatar of users is nonsexual stone statuary
\cite{jarvela2021augmented}. 

\subsubsection{Interactions.} 
\begin{figure}[h]
  \centering
  \includegraphics[width=1\linewidth]{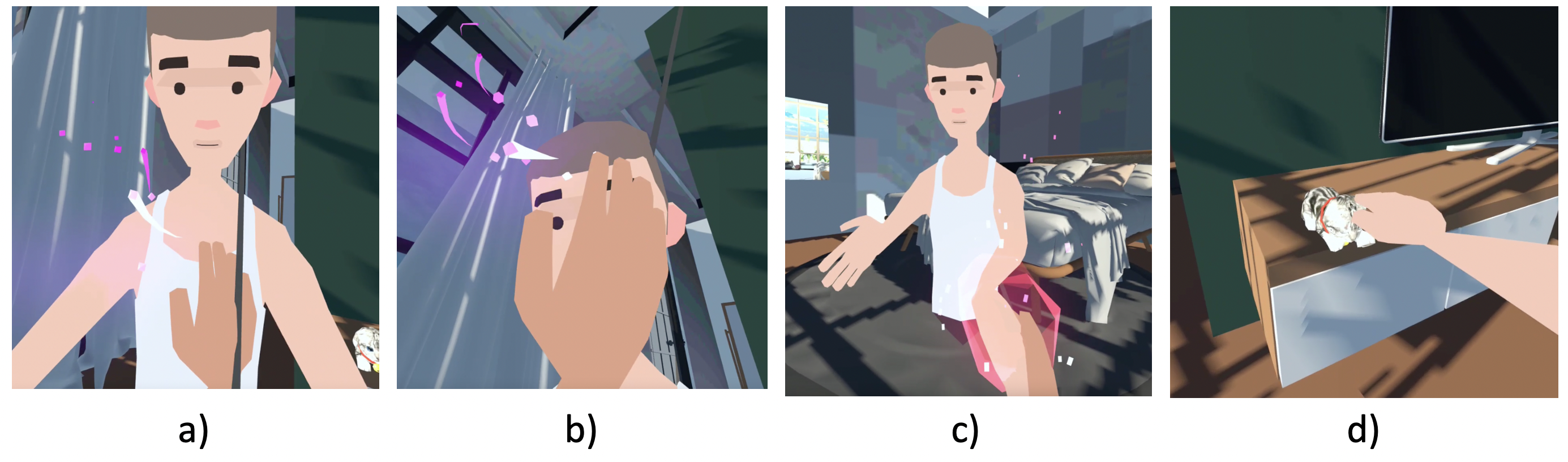}
  \caption{Prototype's interaction and effects: a) fireworks particle effect (FPE) once the contact between the avatar's hand and chest, b) FPE generated by touching the avatar's head with hand, c) love rotation effect when the contact between the avatars' hands. d) avatar's hand interaction with a kitten.}
  \Description{The interaction and effects in the prototype. a), b), c) show the interaction effects between avatars, where a) is the fireworks particle effect generated by the contact between the avatar's hand and chest, b) is the fireworks particle effect generated by touching the avatar's head with hand, c) is the love rotation effect generated by the contact between avatar's hand and hand. d) shows the interaction with the kitten with the avatar's hand.}
  \label{fig:demoInteraction}
\end{figure}

The prototype gives users feedback through special visual effects to enhance the feeling that they are interacting intimately. When the user's avatar hands touch each other, the two hands clasped together will show a heart-spinning effect, and when the user touches another user's head and body with the hands, there will be a short sparking particle effect ( Figure~\ref{fig:demoInteraction}a, b and c). Besides interaction between users, an interaction between the user and the environment was realized. When the user pets the cat, a meow \textasciitilde sound poses very enjoyable experience (Figure~\ref{fig:demoInteraction}d). 

\subsubsection{Meditation experience.}
We included the guided audio, adapted from the one used in the workshop to make it synchronize with the visual content. Besides, we limited users’ mobility and interaction in the VR space after they start meditating to ensure that their meditation is not easily interrupted~\cite{wang2022reducing}. 
After users press the `start meditation' button in the program, the home scene will fix the user's position on the bed, and then play the meditation guide audio to start the meditation for about 10 minutes. The audio will instruct the user to relax in the warm home and enjoy the natural view.

\subsection{Prototype Evaluation}

\subsubsection{Participants}

We recruited eight couple participants (N = 16, 8 male, 8 female) aged 21 - 32 ($M = 25.25$, $SD = 3.04$) from a social network, with a mean familiarity with the VR device of 2.90 ($SD = 1.28$, on a 5-point Likert scale), and all participants self-reported interest in LKM, with a mean familiarity with the meditation of 2.56 ($SD = 0.87$, on a 5-point Likert scale). All participants' visual acuity was normal or corrected to normal. 

\subsubsection{Experiment design}
Each participant would experience both VC LKM and VR LKM sessions for 10 minutes, decided by Latin square ~\cite{grant1948latin}. 
In the VR LKM condition, participants meditate using the developed prototype. In the VC LKM condition, participants meditate using Zoom, a popular video conferencing platform. To control other variables, the conferencing background, the meditation guidance audio, the white noise and the background sound were set the same as in the prototype. The instruction about eye-opening or eye-closing was also kept consistent with the prototype condition, basically following the audio or performing as participants felt comfortable.
Each participant was asked to complete psychological instruments before the meditation, which was seared as the \textit{baseline}. After the first meditation session, the participant needed to complete the instruments again. After a 10-minute break, it’s the second meditation session and psychological assessment. Finally, a brief interview asked for qualitative feedback. During the experience, participants were arranged in two rooms to provide a physically isolated environment. 



\subsubsection{Measurements}
\label{sec:questionnaire}
We collected both participants' responses to psychological instruments and qualitative feedback to evaluate our prototype. 
First, we used three psychological instruments to measure participants' changes in positive and negative emotions, compassion, and the experience of close relationships. 
First, International Positive and Negative Affect Schedule Short Form (PANAS-Short) measured the positive and negative affect with 5 items, respectively, with a 5-point Likert scale for each item~\cite{Panas}. A higher score indicates more positive or negative affect, respectively. 

Second, the compassion aspect was measured by Compassion Motivation and Action Scales (CMAS-self and CMAS-other)~\cite{compassionscale}. For both CMAS-self and CMAS-other, we deleted the items that measure the compassion actions since changes in compassion actions would require a long-term study to observe, and our current lab study focuses on potential short-term effects. The remained items used a 7-point Likert scale. In CMAS-self, 5 items measure the compassion intention and 7 items measure the compassion distress. Higher scores indicate more self-compassion. In CMAS-other, 3 items measure the compassion intention and 3 items measure the compassion distress, with higher scores indicative of more compassion to others. In addition, we adapted words describing other people like \emph{``others''} and \emph{``people''} as \emph{``my partner''} in all items to better fit the close relationship context. 

Third, the experience of close relationship was measured by Experience in Close Relationship Scale - Short Form (ECR-S)~\cite{closerelationship}, in which each item corresponds to a 7-point Likert scale. 6 items measure attachment avoidance, and 6 items measure attachment anxiety. People who score high on either or both of these dimensions are assumed to have an insecure adult attachment orientation.

Regarding qualitative feedback, we mainly asked participants’ feelings for both tools, e.g. the strengths of using the prototype and Zoom, respectively, and participants' willingness to use our prototype for the long term.  

\subsubsection{Statistical analysis methods}
Our study was $1 \times 3$ (1 factor, 3 levels) within-subject design. To analyze the responses to psychological instruments in Section~\ref{sec:questionnaire}, we used the Shapiro-Wilk test to check that the data were normally distributed and the Brown-Forsythe test to confirm that the data were homogeneity of variances, and then applied the parametric analysis method named One-way repeated measures ANOVA analysis.

\subsubsection{Results}
The participants' prototype experience results are discussed in the below two parts. 

\begin{figure}[h]
  \centering
  \includegraphics[width=1\linewidth]{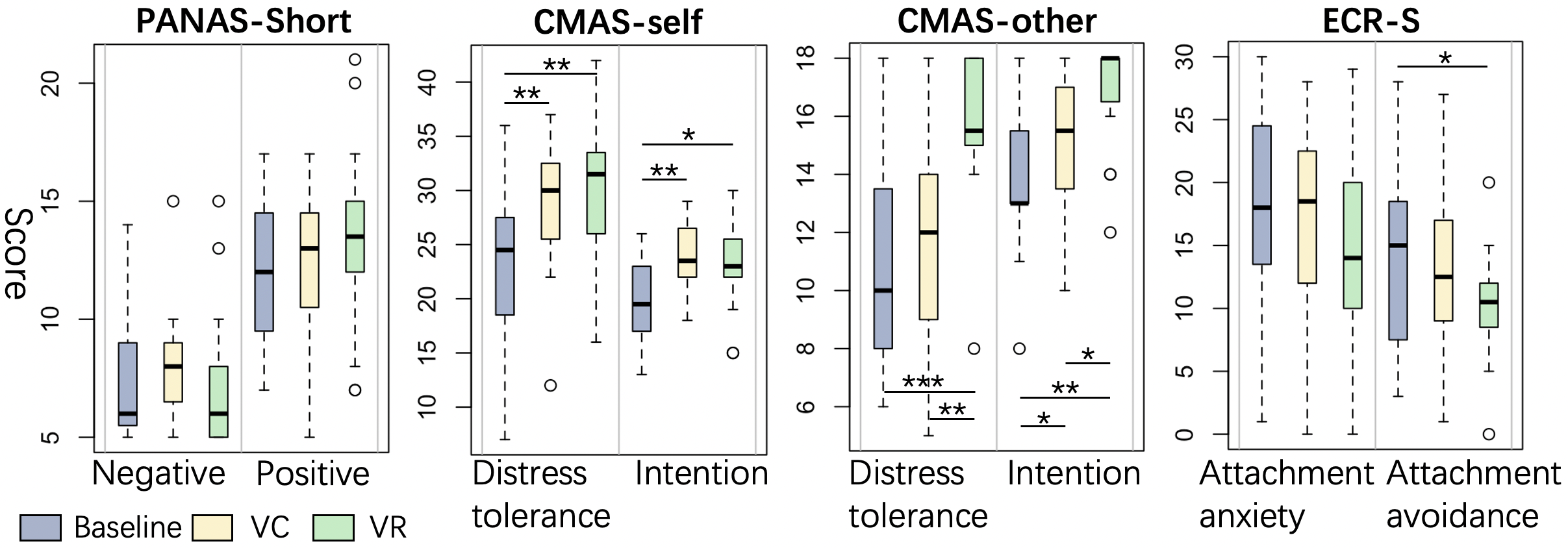}
  \caption{Participants' scores on psychological instruments after Baseline, completion of VC LKM and completion of VR LKM. 
  ($P<.05$(*), $P<.01$(**), $P<.001$(***))}
  \label{fig:PrototypeResult}
\end{figure}

\textbf{(1) Psychological measurement results.}
Figure~\ref{fig:PrototypeResult} depicts the questionnaire results 
before the LKM (Baseline) and after the LKM (VC or VR condition), with details as follows.
\label{sec:Quantitative}

\textit{PANAS-Short.} 
No significant difference exist in \emph{Negative affect} between Baseline ($M = 7.438$, $SD = 2.828$), VC ($M = 8.063$, $SD = 2.435$) and VR ($M = 7.125$, $SD = 3.096$) conditions ($F(2, 30) = .828, p = .447$). However, we can see from the visualization that the average negative effect after participants completed VC LKM was slightly higher than Baseline. After examining the specific items of the PANAS-Short, we discovered that the \emph{Ashamed} item scores for Negative impact increased significantly, to be discussed in Section~\ref{sec:Feedback}. The visualization of the results reveals a trend of slowly increasing \emph{Positive affect} scores from the Baseline ($M = 12.125$, $SD = 3.243$) to VC ($M = 12.313$, $SD = 3.572$) and to VR ($M = 13.500$, $SD = 3.983$) conditions. However, this trend is not statistically significant ($F(2, 30) = 1.108, p = .343$).

\textit{CMAS-self.} The conditions significantly affect the \emph{self-compassion distress tolerance} subscale ($F(2, 30) = 15.405, p<.0001$), with Baseline ($M=22.875$, SD=$8.397$), VC ($M=28.563$, SD=$6.264$) and VR ($M=30.063$, SD=$6.923$), and the \emph{self-compassion intention} subscale ($F(2, 30) = 10.326, p < .0005$) with Baseline ($M=19.688$, SD=$3.683$), VC ($M=24.063$, SD=$3.151$) and VR ($M=23.000$, SD=$4.099$) of the CMAS-self. Post-hoc tests confirm that participants had significantly higher ($p<.01$) \emph{self-compassion distress tolerance} scores after completing VC LKM than Baseline, and after VR LKM also had significantly higher ($p<.01$) \emph{self-compassion distress tolerance} scores than Baseline, but there was no significant difference ($p=.059$) between VC and VR conditions. After VC LKM ($p<.01$) and VR LKM ($p<.05$), participants will have a significantly higher \emph{self-compassion intention} than Baseline, but also not significant compared to VC LKM and VR LKM ($p=.203$). 

\textit{CMAS-other.} The participants' intention to have compassion ($F(2, 30) = 10.175, p<.0005$) for their partner and their availability of tolerance ($F(2, 30) = 10.175, p<.0005$) when their partner was suffering were significantly affected by the LKM method. Post-hoc tests confirm that compared to Baseline ($M=10.813$, SD=$3.563$), participants' scores on the \emph{compassion distress tolerance} subscale of the CMA-other increased significantly after VR LKM ($M=15.813$, SD=$2.562$, $p<.0005$), after VC LKM ($M=11.563$, SD=$3.881$) scores were slightly but not significantly higher ($p=.423$) than Baseline.  Scores on LKM using VR were significantly higher ($p<.01$) than VC. In addition to this, scores on the \emph{compassion intention} subscale also increased significantly after VR LKM ($M=16.813$, SD=$1.870$, $p<.01$) compared to Baseline ($M=13.750$, SD=$2.436$), after VR LKM ($M=14.975$, SD=$2.729$) the scores slightly increase but not significant ($p=.092$), VR LKM affects more ($p<.05$) than VC. 

\textit{ECR-S.} Baseline ($M=18.500$, SD=$7.737$), VC ($M=17.563$, SD=$7.155$), and VR ($M=15.063$, SD=$7.637$) have a significant effect ($F(2, 30) = 4.463, p<.05$) on the \emph{attachment anxiety} scores, however, post hoc analysis did not reveal significant differences between the conditions. The values of the \emph{attachment avoidance} subscales for Baseline ($M=13.813$, SD=$7.250$), VC ($M=12.250$, SD=$6.496$), and VR ($M=10.188$, SD=$4.370$) were significantly different ($F(2, 30) = 5.105, p<.05$). Post hoc analyses showed that participants' \emph{attachment avoidance} values decreased significantly ($p<.05$) after performing VR LKM compared to Baseline.

\textbf{(2) Participants' feedback.}
\label{sec:Feedback}
At the end of the experiment, each couple was interviewed to report their experiences, and their responses were categorized according to the following themes: (1) the advantages and disadvantages of the VR prototype compared with VC as a tool for remote LKM; (2) the willingness to use the prototype as a remote LKM tool for couples in specific scenarios.

\textit{Advantages and disadvantages of VR prototype.}
The VR prototype had the following advantages. All couples (N = 16) agreed that VR is preferable to VC for LKM in remote settings, and the vast majority (N = 12) of participants stated that VR is more immersive and enjoyable than VC, and that a more immersive environment would make them more immersed during LKM. In addition, because VR made them feel as if they were with their partner, they appreciated each other more during LKM. Some participants (N = 8) mentioned that, despite the fact that the avatar was not particularly attractive, they liked the interactive function of the avatar, and the special effects following avatar interaction made them more willing to interact, which was positive feedback. These features were unavailable in VC. Moreover, two participants mentioned that LKM with VR was more beginner-friendly than VC as they require extra effort or higher meditation techniques to focus on LKM facing the distractions under VC conditions, e.g. noise from the earphone. 

VC also had some advantages over VR. Several participants (N = 6) mentioned that they felt facial expressions were important and that they frequently required expressions to communicate. However, the avatar in VR lacks expressions in the current prototype, whereas in VC, the partner's expressions can be accurately viewed, and participants can perceive each other's emotions from the expressions. However, couples (N = 10) also believe that VC's ability to display each other's facial expressions will make them feel awkward and unnatural. And because VC can make them feel as if they are in a meeting or interview, it can cause them to experience some anxiety. This may explain our previous data analysis result, which revealed that, following VC LKM, the average negative emotions of participants increased, as opposed to Baseline.

\textit{Willingness to continue using the VR prototype and other expectations.}
Most participants exhibited great enthusiasm for the VR LKM, and a few (N = 2) even expressed a desire to be long-term contributors to our project using this prototype. However, more participants (N = 10) hoped that if they used the prototype for an extended period, the virtual scenes could be customized, and they would like to create their own space or experience richer virtual scenes. Some couples (N = 4) also desired more realistic interactions, e.g., incorporation of haptic feedback and avatar facial expressions.

\section{Conclusion and Discussion}
This paper examines how long-distance couples can use VR as a tool for LKM to improve intimate relationships. 
By performing LKM with the VR prototype, participants' romantic relationships were enhanced as their \emph{positive affect} increased, \emph{compassion} (both to self and other) significantly increased, and \emph{attachment avoidance} significant decrease. Despite that more long-term studies are needed to explore its long-term effects, our research provides a positive case that VR can be designed to help long-distance couples to practice LKM remotely together, which could potentially be a beneficial intimate social activity for them to calm down and restore the relationship following an unsatisfactory conversation. 
Our prototype evaluation also provided evidence that VR has exclusive advantages for remote LKM practice. Compared with VC LKM, participants had a decrease in \emph{negative affect} and a significant increase in \emph{compassion to others} after practicing VR LKM. The result suggested that VR LKM is more friendly with meditation novices to cultivate compassion, which is consistent with the current study ~\cite{atkinson2013mindfulness}. 
We discuss the Design implication of LKM in VR, as follows.




\textbf{Virtual environment}.
Our findings showed that participants desired a safe environment, including natural elements. This preference is aligned with several environment psychological theories such as attention restore theory~\cite{kaplan2010directed, kaplan1995restorative}, stress reduction theory~\cite{hartig1991restorative, ulrich1991stress},  and prospect-refuge theory~\cite{dosen2016evidence, ruddell1987prospect}. These theories suggested that an environment with a safe cover(refuge), a proper prospect, and fascinating natural scenes can make people restored, manifested in stable emotions, positive affect, lower heart rate, and lower blood pressure. Meditation requires practitioners to calm down first; therefore, designers could refer to these theories to create a relaxed and restorative environment for users. 

Besides the environmental settings like space configuration, we found participants’ preferences for "life", for example, pets or unthreatening critters. They explained that a moving or bleating cat could make them feel the “vitality of life” and further bring them positive emotions. Literature suggests that people like animals because of innate sensitivity and the tendency for biological organisms ~\cite{jacobs2009we}. 

\textbf{Sense of Agency.} 
Participants were highly demanding in customizing the virtual scenarios, with each couple mentioning some precise and specific virtual objects, such as a swing lavender field, bell orchid, blue cat, cinnamon roll, and so on. Other participants would describe scenes with shared memories. Therefore, designers should leave space for users to customize their virtual environment by preparing pre-made components or allowing users to upload individual virtual objects. Besides, LKM can also be well integrated into some existing multi-user VR; for example, Rukangu et al. designed a shared virtual family room for users to conduct daily activities such as chess, golf, and 
 lego-builder~\cite{rukangu2020virtual}. The vision of the virtual family room is to create a framework in which various daily interactions could be added. Therefore, a meditation room or garden could be integrated into the virtual family framework. As mentioned in the workshops, such integration fits with the participants’ desire to own and belong to the space. 

\textbf{Avatars}.
Participants had the desire to see realistic avatars or at least representative avatars based on their appearances. Because in romantic relationships, people’s individual characteristics are important to each other. Besides, partners have a strong desire for the sense of being together. Some studies have already explored user-friendly approaches to creating realistic avatars based on the appearance of users in real life~\cite{saito2016pinscreen, nagano2018pinscreen, li2017pinscreen}. Future research could explore techniques for capturing users' facial expressions, 
allowing users to express emotions more comfortably, which is also mentioned by participants when stating the advantages of ZOOM. 
However, the avatars should be carefully designed, as inappropriate design or representation would have an unexpected impact on users’ compassion cultivation. For example, the uncanny valley effect, see Section ~\ref{sec:fundingInteraction}, may cause negative emotions influencing empathy cultivation, also mentioned in~\cite{jarvela2021augmented}. Therefore, it is important to strike a balance between being representative based on the user’s real appearance and compassion cultivation in LKM practice.  

\textbf{Interaction}.
Interaction is important in VR because virtual character behavior in VR is consistent with how bodies are used in the real world~\cite{freeman2021hugging}, and intimate non-verbal interactions such as hugs and hand-holding by couples in VR can strengthen emotional relationship building, which is also consistent with previous research~\cite{freeman2021hugging,zamanifard2019togetherness}. Multisensory interactions that mimic intimate behaviors can be especially instrumental for LKM VR applications, for example, the technique to realize kissing in VR ~\cite{zhang2021turing}
However, our study found that overly rich interactive content may distract users during meditation (see Section~\ref{sec:avatars}), especially those who are newly exposed to VR. This was also mentioned in a previous study~\cite{prpa2018attending}. Thus, the way of arranging interactive content and meditation progress is a problem for future designers to consider. 



\textbf{Bio-signals \& privacy}.
Couples often prefer not to display their negative moods to each other due to concerns about affecting their partner negatively, opting instead to present their best selves. Negative emotions are typically kept private, whereas positive feelings are shared. Consequently, the ability to observe a partner's bio-signals might not be a positive design feature as it could lead to constant emotional awareness, causing distraction and hindering a meditative state~\cite{lee2022understanding}. Surprisingly, in LKM, privacy was less prioritized, with fewer requests for permissions-related settings, such as visibility controls or interaction permissions~\cite{maloney2020anonymity}. Participants noted that bio-signal results aren't easily controlled, particularly regarding visibility without permission.

\textbf{Limitation and Future Work}
Our participants were from Eastern cultures, and not all of them had rich meditation experiences. 
We did not investigate the potential effects of their backgrounds on their preferences for LKM in VR, such as the length of their long-distance relationships, sexual orientation, and cultural backgrounds. Future work could include more participants with diverse backgrounds. 
To understand the effects of LKM in VR on the relationships and emotional restoration of long-distance couples, more longer-term studies are needed. 


\begin{acks}
This work was supported by the Guangzhou Nansha District Bureau of Science and Technology (under grant number 2022ZD012).
\end{acks}

\bibliographystyle{ACM-Reference-Format}
\balance
\bibliography{mybib.bib}

\end{document}